\documentclass[aps, prb, twocolumn,amsmath,amssymb,floatfix, reprint]{revtex4-2}
\usepackage{graphicx}
\usepackage{soul}
\usepackage{svg}
\usepackage{times}
\usepackage{dcolumn}
\usepackage{hyperref}
\hypersetup{colorlinks, allcolors=blue}
\begin{document}
\title{Orbital currents in lattice multiorbital systems: \\Continuity equation, torques, and RKKY interaction}
\author{Niels Henrik Aase}
\affiliation{\mbox{Center for Quantum Spintronics, Department of Physics, Norwegian University of Science and Technology, NO-7491 Trondheim, Norway}}
\author{Erik Wegner Hodt}
\affiliation{\mbox{Center for Quantum Spintronics, Department of Physics, Norwegian University of Science and Technology, NO-7491 Trondheim, Norway}}
\author{Jacob Linder}
\affiliation{\mbox{Center for Quantum Spintronics, Department of Physics, Norwegian University of Science and Technology, NO-7491 Trondheim, Norway}}
\author{Asle Sudb{\o}}
\email[Corresponding author:]{ asle.sudbo@ntnu.no}
\affiliation{\mbox{Center for Quantum Spintronics, Department of Physics, Norwegian University of Science and Technology, NO-7491 Trondheim, Norway}}

\begin{abstract}
Utilizing the electron orbital degree of freedom in heterostructures is attracting increasing attention due to the possibility of achieving much larger conversion rates between charge and orbital angular momentum flow compared to the intrinsic electron spin. Here, we consider orbital angular momentum currents in a tight-binding multiorbital lattice model and derive their continuity equation. From it, we observe that the current is not conserved and apply similar considerations to the recently discovered altermagnets. We find nonzero orbital torque terms, elucidate their physical mechanism, and show numerically that they contribute to dampening the orbital angular momentum current flowing in multiorbital heterostructures. Moreover, we compute the orbital RKKY interaction and find it exhibits similar characteristics as the orbital angular momentum current mediating it, thus serving as a direct experimental probe of such currents.
\end{abstract}

\maketitle

\textit{Introduction.}
Using the electron's orbital degree of freedom to induce magnetization dynamics has recently garnered considerable attention \cite{Go2017, Go2021, Atencia2024}, establishing the field of orbitronics. Several orbital counterparts to well-known spin effects have been measured, such as the orbital Hall effect \cite{Lee2021, Lyalin2023, Choi2023}, the orbital Edelstein effect \cite{Ding2022, Kim2023, Mendoza-Rodarte2024} and its inverse \cite{ElHamdi2023, Seifert2023}. In many cases, these effects are orders of magnitudes greater than the corresponding spin effects. Moreover, orbitronics, unlike spintronics, does not require spin-orbit coupling (SOC), and orbital effects can be found in light metals such as Ti \cite{Choi2023}.

Orbital angular momentum (OAM) currents can play a pivotal role in memory devices by their ability to control the magnetization in a nearby ferromagnet (FM) \cite{Go2021}. In normal metal (NM)/FM bilayers, the OAM current asserts itself across the interface by applying an orbital torque on the FM. Vice versa, pumping the FM can generate an OAM current inside an NM \cite{Hayashi2024}. The orbital torque is often significantly larger than the conventional spin-orbit torque \cite{Lee2021, Hayashi2023a}. Moreover, it decays far more slowly inside the FM \cite{Bose2023, Moriya2024}. 
The remarkable control asserted over the interplay of OAM currents and magnetism in these experiments is paving the way for orbital-assisted magnetic random-access memory \cite{Gupta2024}.  

Despite the considerable experimental progress made with OAM currents, our understanding of them is still lacking. Fundamental questions pertaining to the conservation of OAM currents remain unanswered. As pointed out in Ref.\ \cite{Atencia2024}, this limits our understanding of crucial aspects in orbitronics, such as information processing, OAM transport and orbital Hall effect, and orbital torque generation. One feature of non-conservation of OAM without a spin counterpart is the emergence of orbital torsion \cite{Han2022}, found using a momentum-space model mainly applied to a two-orbital system. Another is the mechanisms of the angular momentum (both spin and orbital) transfer dynamics in systems with SOC \cite{Haney2010}, viewed through the lens of first-principle calculations \cite{Go2020}.

In this letter, we also consider the obstacle of non-conserved OAM currents. Recently, symmetry constraints have been used to show that orbital moments can relax across a single lattice constant \cite{Urazhdin2023}, and the quantum Liouville equation has been employed to show a strong quantum correction to the OAM current \cite{Liu2024}, as well as the emergence of an intrinsic torque on the OAM in an electric field \cite{Atencia2024a}. We, however, keep to a tight-binding model,
lending a different perspective into OAM currents because we derive the OAM lattice continuity equations in real space, easing the identification of emergent torque terms and their physical interpretation. 
Moreover, we calculate the OAM currents in finite-size FM/NM/FM trilayers. 
To further highlight the difference between conventional spin currents and OAM currents, we also study spin systems with similar symmetries as the multiorbital systems. Specifically, we consider the recently discovered altermagnets \cite{Smejkal2022, Smejkal2022a, Mazin2022, Noda2016,Smejkal2020,Hayami2019, Ahn2019, Yuan2020, Yuan2021} and show that there exists interesting similarities to the behavior of OAM currents.
Finally, owing to the difficulties of directly detecting non-conserved currents, we use our real-space framework to compute the orbital Ruderman–Kittel–Kasuya–Yosida (RKKY) interaction \cite{Kasuya1956,Ruderman1954,Yosida1957}, which is strongly related to the OAM currents, and propose that it gives a less ambiguous experimental fingerprint of OAM currents.

\textit{Microscopic models.}
We will consider two fundamental type of systems: multiorbital systems and systems with spin-dependent hopping. They each have their associated magnetic moment, $\boldsymbol{L}$ and $\boldsymbol{S}$, 
both fulfilling the angular momentum Lie algebra. While $\boldsymbol{L}$ and $\boldsymbol{S}$ belong to different symmetry groups, SO(3) and SU(2), respectively, to ease notation, we use $\boldsymbol{J}$ whenever we consider something that applies to both $\boldsymbol{L}$ and $\boldsymbol{S}$. The angular momentum algebra can then be expressed as $[J_\alpha, J_\beta] = i\varepsilon_{\alpha\beta\gamma} J_\gamma$,
where $\{J_\alpha\}$ are the appropriate generators for the symmetry groups. For SU(2), these are the Pauli matrices $\{\sigma_\alpha\}$ acting on the conventional basis $(\tilde{c}^{S}_n)^{(\dagger)} \equiv \begin{pmatrix}
    c_{n\uparrow}^{(\dagger)} & c_{n\downarrow}^{(\dagger)}
\end{pmatrix}^{\mathrm{T}}$, where $c_{n\uparrow}^{\dagger}$ ($c_{n\uparrow}$) is the creation (annihilation) operator of an electron at site $n$ with spin up. 

\begin{figure}
    \centering
    \includegraphics[width=\linewidth]{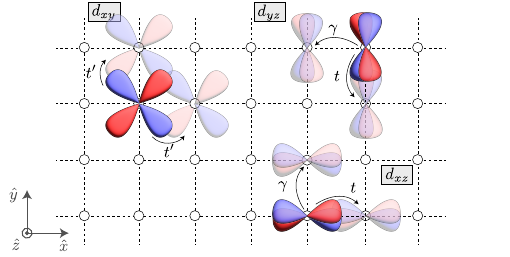}
    \caption{Illustration of the multiorbital tight-binding model. Itinerant $d$ orbitals move on a square lattice, where the hopping strength depends on both the orbital and the direction of movement.}
    \label{fig:lattice_model}
\end{figure}

Unlike spin systems, the basis for multiorbital systems is not canonical. However, recent works on multiorbital superconducting systems \cite{Mercaldo2020, Bours2020, Mercaldo2021, Chirolli2022, Mercaldo2022, Mercaldo2023} developed an effective description of three-orbital systems with nonzero OAM. Both $p$ orbitals \cite{Bours2020} and $d$ orbitals \cite{Mercaldo2020, Mercaldo2021, Mercaldo2022} have been employed. The former are $L=1$ orbitals, while a subset of the latter can form a basis for an \textit{effective} $L=1$ subspace \cite{Mercaldo2020}. We will use $d$ orbitals, specifically $\{d_{yz}, d_{xz}, d_{xy}\}$.
By calculating the matrix elements for the operators $\hat{L}_\alpha = [\hat{\boldsymbol{r}} \times \hat{\boldsymbol{p}}]_\alpha$ and 
the $d$ orbitals,
each component of $\hat{\boldsymbol{L}}$ can be written in matrix form. They are
\begin{equation}
    \tilde{L}_x = \begin{pmatrix}
        0 & 0 & 0 \\
        0 & 0 & i \\
        0 & -i & 0
    \end{pmatrix} \tilde{L}_y = \begin{pmatrix}
        0 & 0 & -i \\
        0 & 0 & 0 \\
        i & 0 & 0
    \end{pmatrix}  \tilde{L}_z = \begin{pmatrix}
        0 & -i & 0 \\
        i & 0 & 0 \\
        0 & 0 & 0
    \end{pmatrix} \label{L_matrices}
\end{equation}
as shown in Ref.\ \cite{Mercaldo2020}, and act on the basis \\
\noindent $\tilde{c}^{L}_n \equiv \begin{pmatrix}
    c_{n, yz} & c_{n, xz} & c_{n, xy}
\end{pmatrix}^{\mathrm{T}}$. Moreover, $\{\tilde{L}_\alpha\}$ are generators of SO(3) obeying the angular momentum algebra. Since SU(2) is the universal covering group of SO(3), so do the $\{\sigma_\alpha\}$. 
We note that the basis in Eq.\ \eqref{L_matrices} does not include spin. This is not an essential simplification since both $L$ and $S$ are good quantum numbers in the absence of SOC. Easing both notation and calculations, we thus omit the spin degree of freedom in the multiorbital systems. In the spin systems, it is still included.

For a two-dimensional square lattice, the relevant point group is $C_{4v}$. Enforcing these symmetries poses significant constraints on the allowed nearest-neighbor hopping processes of the electrons in our multiorbital system. In fact, the only nonzero processes that do not break neither $M_z$ symmetry, nor any of the $C_{4v}$ symmetries, are orbital-preserving processes \cite{Mercaldo2022}. Additionally, fourfold symmetry enforces that $d_{yz}$ orbitals hopping along the $\hat{x}$ and $\hat y$ axis between must be equal to $d_{xz}$ orbitals hopping along the $\hat y$ axis and $\hat x$ axis, respectively. Thus, based on symmetries alone, the most general hopping matrices we can write for the three-orbital model is
\begin{equation}
    \tilde{t}_{n,n\pm e_x}^{L} = \text{Diag}(-\gamma, -t, -t),\; \tilde{t}_{n,n\pm e_y}^{L} = \text{Diag}(-t, -\gamma, -t),
    \label{t_orbit}
\end{equation}
where $e_x$ and $e_y$ are unit vectors in the $x$ and $y$ direction, respectively. $t$, $\gamma$, and $t'$ thus parametrize the strength of the different hopping processes and are all nonnegative.
Figure \ref{fig:lattice_model} depicts the orbitals moving on the lattice and illustrates why, in general, $\gamma<t$ due to the small orbital overlap in the direction of hopping.
Equation \eqref{t_orbit} has no symmetry-lowering terms, rendering it a truly minimal model. Thus, our findings are not rooted in breaking any specific crystal symmetry, making them general. Still, in lower-symmetry models, exotic physics may emerge, such as orbital Rashba effects \cite{Mercaldo2020, Chirolli2022}, that our simpler model cannot capture.
Here, we stress that the movement of the orbitals is intrinsically connected to the underlying crystal structure and generally $\gamma \neq t$. 
As we will see shortly, $\gamma\neq t$ is sufficient to introduce qualitatively new effects, effects that should be present in materials our model may apply to (see next paragraph), regardless of their precise value of $\gamma$.
Conversely, the electron spin plays no role in the kinetics of normal metals, barring SOC.
This difference in how spin and orbital angular momentum relate to spatial symmetries will be important in the following sections.
Lastly, we take $t'=t$ for simplicity, give energies in units of $t$ throughout the paper, and put the lattice constant $a$ and $\hbar$ to unity.

It is instructive to consider materials the multiorbital tight-binding model might apply to. In elemental materials with tetragonal or cubic crystal structures, symmetry enforces direction dependency on the hopping \cite{Mercaldo2020}, such that the hopping matrices in Eq.\ \eqref{t_orbit} can describe elements such as, \textit{e.g.}, Ti, Ni, and Nb (Al), which are elements with a prominent $d$ ($p$) orbital character.
We also highlight that the angular momentum of $d$ orbitals has been elucidated in recent experimental studies of OAM textures in topological materials such as CoSi \cite{Brinkman2024} and TaAs \cite{Figgemeier2024} using dichroism and angle-resolved photoemission spectroscopy. Specifically, in Ref.\ \cite{Brinkman2024}, it is exactly the eigenstate of $\tilde{L}_x$, $d_{xz} + i d_{xy}$ that is measured. The model we employ was also recently applied to the surface of Sr$_2$RuO$_4$ in a study of spin-orbital chiral surface currents \cite{Mazzola2024}.

We also consider the simpler kinetics in spin systems.
In altermagnets, the hopping can be described by an effective model \cite{Ouassou2023}
\begin{equation}
    \tilde{t}_{n,n\pm e_x}^{S, \mathrm{AM}} = -t \mathbb{I} + w\sigma_z \quad \tilde{t}_{n,n\pm e_y}^{S, \mathrm{AM}} = -t \mathbb{I} - w\sigma_z,
    \label{t_AM}
\end{equation}
where $w$ is the spin anisotropy. We include altermagnets in our study because of their similarities to the multiorbital systems. Namely, in both Eqs.\ \eqref{t_orbit} and \eqref{t_AM}, there is an intimate link between the spatial symmetry of the crystal and the angular momentum of the electrons.

Including a chemical potential $\mu$, the Hamiltonian for both the multiorbital and spin systems can be expressed concisely as
\begin{equation}
    H_{\mathrm{kin}} + H_\mu = \frac{1}{2}\sum_{\langle n,m\rangle} (\tilde{c}_n^J)^\dagger \tilde{t}_{nm}^{J} \tilde{c}_m^J + \mathrm{h.c.} - \mu\sum_n (\tilde{c}_n^J)^\dagger \tilde{c}_{n}^J,
    \label{H_standard}
\end{equation}
where the first sum is over nearest neighbors. Moreover, since both $\boldsymbol{L}$ and $\boldsymbol{S}$ are magnetic moments, they couple to magnetic fields. Thus, introducing $\tilde{\boldsymbol{J}} = \begin{pmatrix}
    \tilde{J_x} & \tilde{J_y} & \tilde{J_z}
\end{pmatrix}^\mathrm{T}$, their coupling to a ferromagnet with magnetization $\boldsymbol{m}$ can be written as $H_{\mathrm{FM}} = -\sum_{n \in \mathrm{FM}} \boldsymbol{m} \cdot (\tilde{c}^J_n)^\dagger\tilde{\boldsymbol{J}}\tilde{c}^J_n$
regardless of whether the magnetization stems from intrinsic electron spin, such as in Ni and Fe, or from the orbital degree of freedom, dominant in some moiré heterostructures \cite{Tschirhart2021}. Inside the FMs, the hopping matrix $\tilde{t}_{nm}^{J, \mathrm{FM}} = -t\mathbb{I}$ is taken to be isotropic.
We will consider bilayer and trilayer structures comprised of the above systems, \textit{e.g.}, a multiorbital NM, proximitized to two ferromagnets with magnetizations $\boldsymbol{m}_{\mathrm{L}}$ and $\boldsymbol{m}_{\mathrm{R}}$.
The entire heterostructure is then described with $H_{\mathrm{kin}}$, $H_\mu$ and $H_{\mathrm{FM}}$, and by using the appropriate hopping matrix and magnetizations for each of the subsystems.

\textit{Currents and torques.}
Systems with SOC can host equilibrium spin currents in the absence of external fields \cite{Rashba2003}. Moreover, these systems host non-conserved currents due to emergent torque terms acting on the spin, viewed originally as an obstacle for the conventional definition of spin current\cite{Shi2006}, and discussed further in Refs. \cite{Sonin2007, Tokatly2008}.
We encounter a similar situation for the orbital current. 
Analogous to the definition of the spin current, we define the orbital current as the operator coupling linearly to the SO(3) gauge potential of the OAM. This has the appealing property that, as we will see shortly, the orbital current is conserved in a system with isotropic hopping, similar to how a spin current is conserved in the absence of SOC. Thus, following the derivation and subsequent definition of the spin current outlined in Ref.\ \cite{Fujimoto2021}, we replace the SU(2) generators with the SO(3) ones where needed. 
The general expression for both spin and orbital current at site $n$ with polarization $\alpha$ moving in the $k$ direction then becomes
\begin{align}
    j_{n,\alpha,k}^J &= \frac{ia^2}{2}(\tilde{c}_n^J)^\dagger \tilde{J}_\alpha \tilde{t}_{n, n+e_k}^J \frac{\tilde{c}^J_{n+e_k} - \tilde{c}^J_{n-e_k}}{2a} + \text{h.c.},
\label{gen_eq_for_currents}
\end{align}
where we briefly reinstate $a$ to highlight the lattice derivative.

To further analyze the currents, we derive their associated continuity equations. We use the Heisenberg equation $\dot{\hat{\eta}}=i[H, \hat{\eta}]$ where $\hat{\eta}$ is a general operator. For an angular momentum operator $\hat{J}_{\alpha,n} = (\tilde{c}^J_n)^\dagger\tilde{J}_\alpha\tilde{c}^J_n$, from straightforward calculations of its commutator with $H$, it then follows that its time dependence is given by
\begin{align}
    \dot{\hat{J}}_{\alpha,n} = \frac{i}{2} \sum_\delta \left( \tilde{c}^{\dagger}_{n+\delta} \tilde{t}_{n+\delta,n} \tilde{J}_\alpha\tilde{c}_n - \tilde{c}^\dagger_n \tilde{J}_\alpha\tilde{t}_{n,n+\delta}\tilde{c}_{n+\delta}\right),
    \label{heisenberg_output}
\end{align}
where the sum is over all nearest neighbors. Carrying out this sum, Eq.\ \eqref{heisenberg_output} yields two types of terms, those which can be written as a divergence of the current in Eq.\ \eqref{gen_eq_for_currents} (with lattice derivatives) and those which cannot. Collecting the latter and defining them as $\tau$, we can rewrite Eq.\ \eqref{heisenberg_output} as a continuity equation
\begin{align}
    \dot{\hat{J}}_{\alpha,n} + \boldsymbol{\nabla} \cdot \boldsymbol{j}^{J}_\alpha = \sum_k \tau_{\alpha, n, k}^J,
    \label{cont_eq_general}
\end{align}
where $\tau_{\alpha, n, k}^J$ acts like a torque term and is given by
\begin{align}
    \tau_{\alpha, n, k}^J = i \tilde{c}_n^\dagger [\tilde{t}_{n+e_k,n},\tilde{J}_\alpha] \tilde{c}_n -\frac{i}{2} \frac{\partial \tilde{c}_{n}^\dagger}{\partial k} [\tilde{t}_{n+e_k, n}, \tilde{J}_\alpha] \frac{\partial \tilde{c}_{n}}{\partial k}.
    \label{tau_general}
\end{align}

Equations \eqref{cont_eq_general} and \eqref{tau_general} constitute an important result: If $\tilde{J}_\alpha$ commutes with the hopping matrix $\tilde{t}^J$, all torques are zero and the angular momentum current is conserved. This is always the case with isotropic hopping. In several systems, however, the hopping matrix is inherently anisotropic, as discussed in the previous section, thus allowing nonzero torque terms and non-conserved currents. 

To elucidate the physics of Eqs.\ \eqref{cont_eq_general} and \eqref{tau_general}, we consider the multiorbital system with generators and hopping matrices given in Eq.\ \eqref{L_matrices} and \eqref{t_orbit}, respectively, calculate the commutator in Eq.\ \eqref{tau_general}, and obtain
\begin{subequations}
    \label{taus_explicit_orbital}
        \begin{align}
            \tau_x^L &= -(t - \gamma) \tilde{c}_n^\dagger|\tilde{L}_x|\tilde{c}_n + \frac{1}{2} (t - \gamma) \frac{\partial \tilde{c}_{n}^\dagger}{\partial y} |\tilde{L}_x| \frac{\partial \tilde{c}_{n}}{\partial y} \\
            \tau_y^L &= (t - \gamma) \tilde{c}_n^\dagger|\tilde{L}_y|\tilde{c}_n - \frac{1}{2} (t - \gamma) \frac{\partial \tilde{c}_{n}^\dagger}{\partial x} |\tilde{L}_y| \frac{\partial \tilde{c}_{n}}{\partial x} \\
            \tau_z^L &= \frac{1}{2} (t - \gamma) \big( \frac{\partial \tilde{c}_{n}^\dagger}{\partial x} |\tilde{L}_z| \frac{\partial \tilde{c}_{n}^\dagger}{\partial x} - \frac{\partial \tilde{c}_{n}^\dagger}{\partial y} |\tilde{L}_z| \frac{\partial \tilde{c}_{n}}{\partial y}\big).
        \end{align}
\end{subequations}
Here, we have introduced $|\tilde{L}_\alpha|$, defined as the element-wise absolute value of $\tilde{L}_\alpha$. A similar operator is defined as the orbital angular position operator in Ref.\ \cite{Han2022}. Applying $|\tilde{L}_\alpha|$ on a site with an orbital linear combination with nonzero OAM in the $\alpha$ direction changes the sign of the OAM. All terms in Eq.\ \eqref{taus_explicit_orbital} scale with $(t-\gamma)$, which is the effective multiorbital-hopping anisotropy parameter. This can be understood physically. The eigenstates with nonzero OAM consist of two orbitals with a complex phase difference, so as the OAM current moves the state from one site to another, the hopping anisotropy breaks the state, resulting in a loss of OAM. However, the OAM can not vanish and is therefore transferred to the lattice.
If the anisotropy increases, so does the loss of angular momentum, in turn explaining why $\tau^L_\alpha$ scales with $(t-\gamma)$.

Equation \eqref{tau_general} can also be used to find the nonzero spin-torque terms in spin systems. 
Specifically, in altermagnets, $\tau^{\mathrm{AM}}_\alpha$ are similar to those of Eq.\ \eqref{taus_explicit_orbital}, with the Pauli matrices replacing $|\tilde{L}_\alpha|$.
$\tau^{\mathrm{AM}}$ are easily found by calculating $[\tilde{t}^{\mathrm{AM}}_{n, n+e_k}, \sigma_\alpha]$, but here we only note that $\tau_z^{\mathrm{AM}}$ is zero, while $\tau_x^{\mathrm{AM}}$ and $\tau_y^{\mathrm{AM}}$ are nonzero. Thus, the component of the spin current polarized in the $z$ direction is conserved, in contrast to the OAM current where none of the components are. We note that the continuity equation for the OAM current displays a rich behavior with regard to torque terms despite the absence of symmetry-lowering terms like orbital Rashba coupling, unlike spin currents.

\textit{Orbital currents in FM/NM/FM junctions.}
While the physics of the source terms in Eq.\ \eqref{tau_general} are clear at a local level, it is not obvious how they affect the current across a larger system. 
Therefore, we consider a multiorbital FM/NM/FM trilayer.
The entire system has $N_x$ columns and $N_y$ rows. Each FM spans $N_{\mathrm{FM}}$ columns.
In the $x$ direction, we employ hard boundary conditions, mimicking the geometry of a real trilayer. However, in the $y$ direction, we use periodic conditions (PBC). Since the system is translationally invariant in the vertical direction, we can then partially diagonalize the system by introducing momentum operators in the $y$ direction, $c_n\rightarrow c_{n_x, k_y}$. This is straightforward and reduces the complexity of diagonalizing the system, going from a $3N_xN_y$ matrix to a block-diagonal matrix with $N_y$ blocks, each with size $3N_x$. These blocks can be diagonalized simultaneously, allowing us to study far larger systems.

\begin{figure}
    \centering
    \includegraphics[keepaspectratio, width=\linewidth]{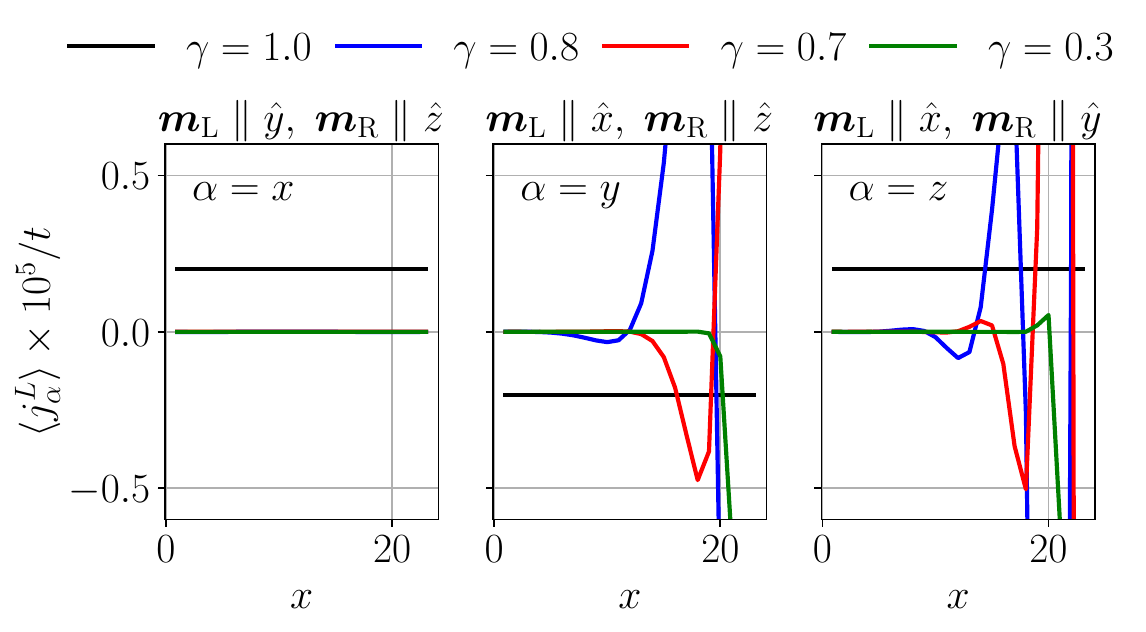}
    \caption{The nonzero OAM current $\langle j^L_\alpha\rangle$ as a function of horizontal position $x$ in FM/NM/FM trilayers. The direction of magnetization $\boldsymbol{m}$ in the FMs is different in each panel, otherwise the system parameters are the same. They are $N_x=25$, $N_y = 800$, $N_{\mathrm{FM}} = 1$, $|\boldsymbol{m}_{\mathrm{L/R}}| = 1$, $T=0.01$, $\mu=-0.2$.
    }
    \label{fig:current}
\end{figure}

We numerically diagonalize the system and subsequently use the $3N_xN_y$ eigenvectors and eigenvalues to determine the quasiparticles $\{\gamma_n\}$ of the system, as well as their energy $\{E_n\}$. The expectation value of a general bilinear operators is determined as follows: $
    \langle c_{i}^\dagger c_{j} \rangle = \sum_{n} U_{in}^* U_{jn} \langle \gamma_n^\dagger \gamma_n \rangle =  \sum_{n} U_{in}^* U_{jn} n_\mathrm{F}(E_n),$
where we suppressed the orbital indices for notational simplicity. Here, $n_\mathrm{F}(x)$ is the Fermi-Dirac distribution at temperature $T$, and $U$ is the modal matrix of the Hamiltonian. 
Since we are considering thermal equilibrium, the time derivative in Eq.\ \eqref{cont_eq_general} is zero. Moreover, because we are in equilibrium, we find that the OAM current in the system is an exchange OAM current, $j(\boldsymbol{r}) \propto \boldsymbol{L}(\boldsymbol{r}) \times \nabla \boldsymbol{L}(\boldsymbol{r})$, generalizing the equivalent spin expression \cite{Maekawa2023}.

In Fig.\ \ref{fig:current}, we consider different magnetization directions in the FM and systems with differing $\gamma$ and calculate the emergent OAM currents.
We first briefly consider the limiting case of $\gamma=t$. The OAM current then behaves in the same manner as a spin current in a conventional FM/NM/FM trilayer: it is conserved and finite (supposing there is a nonzero misalignment between the two FMs), and its direction of polarization is given by $\boldsymbol{m}_\mathrm{L} \times \boldsymbol{m}_\mathrm{R}$. The OAM current is conserved since the torque terms in Eq.\ \eqref{taus_explicit_orbital} are all zero. 

With $\gamma\neq t$ in Fig.\ \ref{fig:current}, $j^L_\alpha$ is no longer conserved. In contrast to spin FM/NM/FM systems, the OAM current not only depends on the relative alignment of the magnetizations but also their absolute directions, creating an orbital angular momentum anisotropy. The OAM current gradually decays as it moves away from the right FM and $j_L^x$ is orders of magnitudes smaller than $j_L^y$ and $j_L^z$.
We can understand this from the expression for exchange OAM current. The emergence of the OAM current is contingent on the orbital magnetization from the FMs permeating into the NM. For $\alpha=y,z$, the magnetization of the left FM is polarized in the $x$ direction and thus consists of $d_{xz}$ and $d_{xy}$ orbitals. These are mobile in the $x$ direction with $t_x = t$ and can thus move far into the NM, carrying with them a finite orbital magnetization. However, when the right FM has an orbital magnetization in the $y$ or $z$ direction, it consists of the less horizontally mobile $d_{yz}$ orbital, therefore not penetrating as far into the NM as the magnetization from the left FM, causing the OAM current to be at its largest close to the right FM.
The poor horizontal mobility of the $d_{yz}$ orbitals also explains why $j_L^x$ diminishes to such an extent in Fig.\ \ref{fig:current}, even for modest anisotropy $\gamma=0.8$: $L_y$ dies off rapidly from the left, while $L_z$ dies rapidly off from the right, leaving no area where they are both of significant size, which is required to have a sizable $j_L^x$. Finally, from Fig.\ \ref{fig:current}, we 
see that the current is dampened quicker with decreasing $\gamma$, in accordance with our analytical results above, demonstrating that the nature of OAM transport precludes a large current across the trilayer considered here. In the extreme case of $\gamma=0$, we obtain $j^L_\alpha=0$, consistent with the results in Ref.\ \cite{Urazhdin2023} where the orbital moments relax across a single lattice constant. 

In Fig.\ \ref{fig:current} and throughout the paper, we use $T=0.01k_\mathrm{B}t$, which, for $t=1$ eV corresponds to $\approx 116$ K. Lowering $T$ beyond $0.01t$ slightly increases, in general, the magnitude of the quantities we consider. Conversely, increasing $T$ diminishes them.
The question of how introducing disorder affects $j_{L}^\alpha$ has also been raised \cite{Atencia2024}. Here, we only note that our model can account for disorder by introducing a random site-dependent term $\epsilon_n$ to the chemical potential $\mu\rightarrow \mu_n=\mu + \epsilon_n$, but we leave this question for future investigations.
Lastly, we note that varying $\mu$ and $|\boldsymbol{m}_{\mathrm{L/R}}|$ changes the magnitude and sign of $j_{L}^\alpha$ at $\gamma=t$, just like in the well-known spin-current case.
Going to $\gamma<t$,
the dampening and oscillation of $j_{L}^\alpha$ persists, $\mu$ and $|\boldsymbol{m}_{\mathrm{L/R}}|$ do not introduce any new effects on their own.

\textit{Orbital RKKY interaction.}
Since the OAM currents are not conserved in general and difficult to measure directly, we compute a different physical observable that is routinely measured while still directly dependent on the OAM currents, in order for us to make experimental contact with our predictions. We compute the RKKY interaction between two orbitally polarized ferromagnets, mediated by the OAM current, and show how it acquires properties that are different from the previously studied RKKY interaction between spins.
From the free energy of the system $F = -\frac{1}{\beta} \sum_n \ln(1 +\mathrm{e}^{-\beta E_n}),$
we determine the induced exchange coupling $J$ between the FMs by finding the free-energy difference between a system with antiparallel FMs $F_{\mathrm{AP}}$, and parallel FMs $F_{\mathrm{P}}$ \cite{Bruno1995}.

\begin{figure}[t!]
    \centering
    \includegraphics[keepaspectratio, width=\linewidth]{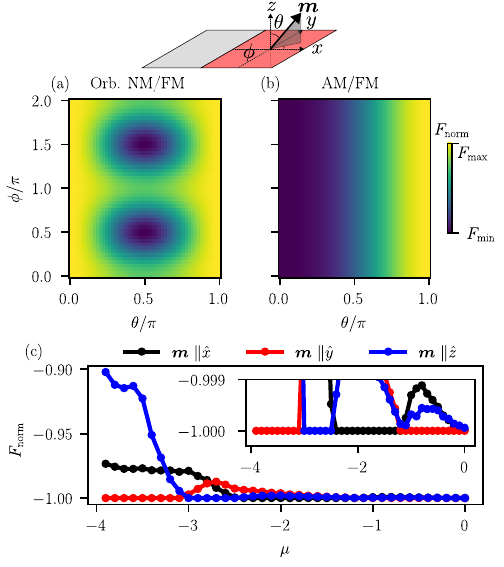}
    \caption{The normalized free energy $F_{\mathrm{norm}}$ for different bilayer systems. The geometry of the bilayers is illustrated at the top, with an FM marked in red. In Figs.\ (a) and (b), $F_{\mathrm{norm}}$ is plotted as a function of the polar angle $\theta$ and azimuthal angle $\phi$ of the magnetization in the FM $\boldsymbol{m}$, while in (c), it is a function of $\mu$ for different directions of $\boldsymbol{m}$, where the inset shows the lowest values of $F_{\mathrm{norm}}$. In (a) and (c), we consider a multiorbital normal metal adjacent to an FM with hopping anisotropy $\gamma=0.3$. In (b), we consider an AM/FM bilayer, where the spin anisotropy in the altermagnet is $w=0.4$. The remaining system parameters are the same as in Fig.\ \ref{fig:current} and $T=0.01$.}
    \label{fig:fNorm_combined_fig}
\end{figure}

Before proceeding further, we must consider an effect not present in conventional spin RKKY interactions. Namely, there exists a direct magnetic proximity coupling in multiorbital NM/FM bilayer systems, where the hopping anisotropy in the NM affects the preferred magnetization direction of the FM despite the fact that no orbital polarization exists in the NM by itself, in total nor momentum resolved. We show this in Fig.\ \ref{fig:fNorm_combined_fig} (a), plotting the normalized free energy as a function of the azimuthal angle $\phi$ and polar angle $\theta$ of the magnetization in the FM.
The free energy minimizes with the magnetization along the $y$ axis. It is degenerate at $\pm \boldsymbol{y}$, in stark contrast to systems with isotropic hopping such as spin NM/FM bilayers. There, the SU(2) symmetry of the Hamiltonian leaves the free energy invariant under rotation of the magnetization. In Fig.\ \ref{fig:fNorm_combined_fig} (a), the corresponding SO(3) symmetry is broken since $\gamma\neq t$ in the NM, allowing an angular dependence in the free energy, as seen in Fig.\ \ref{fig:fNorm_combined_fig} (a).

Interestingly, it is possible to identify similar behavior of the proximity coupling in Fig.\ \ref{fig:fNorm_combined_fig} (a) in spin systems where SU(2) symmetry is broken, but no net spin polarization exists. Specifically, we consider an AM/FM bilayer and plot its normalized free energy in Fig.\ \ref{fig:fNorm_combined_fig} (b).
However, unlike the multiorbital case in Fig.\ \ref{fig:fNorm_combined_fig} (a), the energy minimum is not degenerate; it is at $\boldsymbol{z}$ in Fig.\ \ref{fig:fNorm_combined_fig} (b).
We lastly note an additional fundamental difference between the multiorbital and spin proximity coupling. The degenerate minima along $\pm\boldsymbol{y}$ in Fig.\ \ref{fig:fNorm_combined_fig} (a) can also occur at $\pm\boldsymbol{x}$ or $\pm \boldsymbol{z}$ if $\mu$ is changed, supposing that $\gamma\neq t$, as shown in Fig.\ \ref{fig:fNorm_combined_fig} (c). Figure \ref{fig:fNorm_combined_fig} (c) thus suggests that the direction of magnetization in the FM can be controlled by changing the chemical potential in a multiorbital NM, either through doping or electric gating in the low-dimensional case.

\begin{figure}[t!]
    \centering
    \includegraphics[keepaspectratio, width=0.95\linewidth]{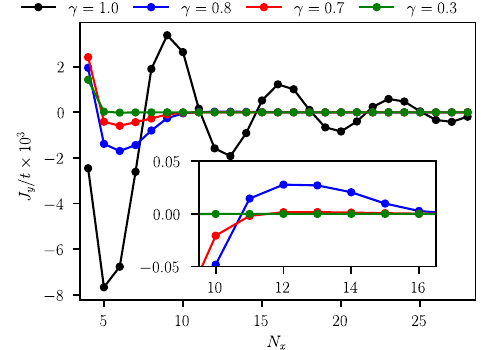}
    \caption{The induced exchange interaction in the $y$ direction $J_y$, between two orbital ferromagnets in an FM/NM/FM multiorbital trilayer as a function of the system length $N_x$. The FMs are polarized along the $y$ axis, and the system parameters are otherwise the same as in Fig.\ \ref{fig:current}. \textit{Inset:} zoom in close to a sign-change of $J_y$ with $\gamma\neq t$.}
    \label{fig:RKKY_func_of_gamma_J_y}
\end{figure}

To break the degeneracy at $\pm \boldsymbol{y}$ in Fig.\ \ref{fig:fNorm_combined_fig} (a), we again consider an FM/NM/FM trilayer, where we fix the magnetization in the left FM in the positive $y$ direction. We then consider the cases of $\boldsymbol{m}_{\mathrm{R}}= \boldsymbol{y}$ and $\boldsymbol{m}_{\mathrm{R}}= -\boldsymbol{y}$, calculating $F_\mathrm{P}$ and $F_{\mathrm{AP}}$ separately, and thus determine $J\equiv F_{\mathrm{AP}} - F_\mathrm{P}$. Since we consider specifically the $y$ direction of polarization, we add a subscript to $J_y$. Then, $J_y$ is calculated for different values of $N_x$, with different values of $\gamma$, to observe the effect of the hopping anisotropy on the RKKY interaction. These results are in Fig.\ \ref{fig:RKKY_func_of_gamma_J_y}. First, we highlight the curve with $\gamma=t$. It exhibits conventional spin RKKY behavior, oscillating and decaying as a function of the distance between the FMs.
The similarity between the spin RKKY and orbital RKKY interaction is expected when $\gamma=t$ since the torque terms in Eq.\ \eqref{cont_eq_general} in both cases are zero. Thus, the spin and orbital currents mediating the RKKY interaction obey the same continuity equation.

The more intriguing aspect of Fig.\ \ref{fig:RKKY_func_of_gamma_J_y} occurs when $\gamma \neq t$. For moderate values of $\gamma$, we can still observe the RKKY oscillations until $N_x=9$. In this regime, $J_z$ is of the order $10^{-3}t$, corresponding to a few meV. From the inset in Fig.\ \ref{fig:RKKY_func_of_gamma_J_y}, we observe that the oscillations in $J_y$ persist for larger values of $N_x$, but it is dampened rapidly. The dampening increases with hopping anisotropy; for $\gamma = 0.3$, $J_y$ is significantly suppressed already at small distances. This is consistent with the behavior of the OAM currents in Fig.\ \ref{fig:current} because the suppression of the OAM currents mediating the RKKY interaction will also affect the interaction itself.
While the RKKY coupling provides a direct experimental probe for OAM currents and the torque terms leading to their non-conservation, the rapid suppression of $J_y$ in Fig.\ \ref{fig:RKKY_func_of_gamma_J_y}, even at modest values for $\gamma$, suggests that short junctions should be considered experimentally since they produce a larger and more easily detectable exchange interaction.

\textit{Concluding remarks.}
Using a tight-binding multiorbital lattice model, we derive the continuity equation for the OAM currents and observe that the OAM is not conserved, with nonzero torque terms acting on the currents.
To elucidate the effect of the torque terms, we consider an FM/NM/FM trilayer numerically and find that the OAM current dampens across the system.
Moreover, to overcome the obstacle of non-conserved OAM currents and their lack of a direct experimental fingerprint, we also compute the orbital RKKY interaction between two FMs in FM/NM/FM trilayers, which is a measurable quantity. We find that the orbital RKKY interaction decays more quickly than its spin counterpart. Our results thus show several orbital effects without a direct analogy in conventional spin systems.

\textit{Acknowledgements.} This work was supported by the Research Council of Norway (RCN) through its Centres of Excellence funding scheme, Project No. 262633, "QuSpin", as well as RCN Project No. 323766.

\bibliography{main.bib}
\end{document}